%% file: V643_Ori_Final.tex
\documentclass{aa}
\usepackage{graphicx}
\usepackage{amsmath}
\usepackage{float}
\usepackage{natbib}
\newcommand\kms{\ifmmode{\rm km\thinspace s^{-1}}\else km\thinspace s$^{-1}$\fi}
\newcommand\vstar{V643~Ori}
\newcommand\mn{MNRAS}

\begin{document} 

\title{The detached, evolved post-mass-exchange binary \\ V643 Orionis}

\author{J.\ Andersen\inst{1,2}
\and
        G.\ Torres\inst{3}
\and
        J.\ V.\ Clausen\inst{1}\fnmsep\thanks{Deceased June 5, 2011.}
}

\institute{The Niels Bohr Institute, University of Copenhagen, Blegdamsvej 17, DK-2100 Copenhagen, Denmark\\
             \email{ja@nbi.ku.dk}
    \and
Stellar Astrophysics Centre, Department of Physics and Astronomy, Aarhus University, DK-8000 Aarhus C, Denmark             
    \and
Center for Astrophysics \textbar\ Harvard \& Smithsonian, Cambridge, MA 02138, USA
\email{gtorres@cfa.harvard.edu}
}

\date{Received; accepted; } 

\abstract
%
%
{One of the greatest uncertainties in modelling the mass-exchange phases during the evolution of a binary system is the amount of mass and angular momentum that has been lost from the system. In order to constrain this problem, a favourable, evolved and detached real binary system is valuable as an example of the end result of this process.}
{We study the 52-day post-mass-exchange eclipsing binary \vstar\ from complete $uvby$ light curves and high-resolution spectra. \vstar\ is double-lined and shows total primary eclipses. The orbit is accurately circular and the rotation of both stars synchronised with the orbit, but the photometry from a single year (1993) shows signs of weak spot activity (0.02 mag) around the primary eclipse.}
{We determine accurate masses of 3.3 and 1.9~$M_{\sun}$ from the spectroscopic orbit and solve the four light curves to determine radii of 16 and 21~$R_{\sun}$, using the Wilson-Devinney photometric code. The rotational velocities from the cross-correlation profiles agree well with those computed from the known radii and orbital parameters. All observable parameters are thus very precisely determined, but the masses and radii of \vstar\ are incompatible with undisturbed post-main-sequence evolution.}
{We have attempted to simulate the past evolutionary history of \vstar\ under both conservative and non-conservative Case~B mass transfer scenarios. In the non-conservative case we varied the amounts of mass and angular momentum loss needed to arrive at the present masses in a circular 52-day orbit, keeping the two stars detached and synchronized as now observed, but without following the evolution of other stellar properties in any detail. Multiple possible solutions were found. Further attempts were made using both the BSE formalism and the binary MESA code in order to track stellar evolution more closely, and make use of the measured radii and temperatures as important additional constraints. Those efforts did not yield satisfactory solutions, possibly due to limitations in handling mass transfer in evolved stars such as these.
We remain hopeful that future theoreticians can more fully model the system under realistic conditions.}
 

\titlerunning{\vstar} 

\maketitle

\section{Introduction}
\label{sec:introduction}

The 52.4-day G-K binary star V643~Orionis (HDE~294651) was first studied in depth by \cite{Imbert:1987}, who observed a double-lined spectroscopic orbit with the photoelectric radial-velocity scanner CORAVEL \citep{Baranne:1979} on the 1\,m Swiss telescope at Observatoire de Haute Provence and the 1.5\,m Danish telescope at ESO, La Silla, Chile. He got all main properties of the present system essentially right and found that both components are detached, synchronously rotating giants, but failed to notice the significant fact that the primary and secondary radii were in the opposite ratio of the masses and luminosities. 

Since then, we have obtained complete, well-covered $uvby$ light curves and even more precise radial-velocity curves, and thoroughly analysed them to get a complete, detailed picture of the system in its present state. This paper documents our observations, analysis, and results on the system. In this way, we hoped to gain some insight into the physical processes that shape the evolution of interacting binaries in general, notably about the processes of mass and angular momentum loss that determine the final outcome. Our original goal, to provide a unique description of the evolution of \vstar\ from the initial configuration to the final state, turned out not to be quite possible, but we eventually came close. 

\section{Spectroscopy}
\label{sec:spectroscopy}

\subsection{Observations and associated results}

Spectroscopic observations of \vstar\ were made at the Center for Astrophysics | Harvard \& Smithsonian (CfA) between October of 1987 and February of 1989 with an echelle spectrograph ($R \approx 35,000$) attached to the 1.5\,m Wyeth reflector at the Oak Ridge Observatory (now closed) in the town of Harvard, Massachusetts. A single order was recorded with an intensified photon-counting Reticon detector providing 45~\AA\ of wavelength coverage centered on the \ion{Mg}{1}~b triplet near 5187~\AA. A total of 44 usable spectra were collected with signal-to-noise ratios ranging from 13 to 24 per resolution element of 8.5~\kms. The velocity zero point was monitored by means of nightly sky exposures taken at dusk and dawn \citep[see][]{Latham:1992}.

\input ntable1.tex

Radial velocities (RVs) for the two components were measured as
described by \cite{Torres:2015} with the two-dimensional
cross-correlation algorithm TODCOR \citep{Zucker:1994}, using
templates taken from a pre-computed library of synthetic spectra based
on model atmospheres by R.\ L.\ Kurucz for solar metallicity
\citep[see][]{Nordstrom:1994, Latham:2002}.  The optimal template for
each star was found by running grids of cross-correlations over a wide
range of effective temperatures ($T_{\rm eff}$) and rotational
broadenings ($v \sin i$) following \cite{Torres:2002}, and selecting
the combination giving the highest cross-correlation value averaged
over all observations, with weights set by the strength of each
exposure. In this way we estimated temperatures of $5210 \pm 150$~K
and $4520 \pm 150$~K for the primary (the more massive star) and
secondary, and projected rotational velocities of $18 \pm 2$~\kms\ and
$23 \pm 3$~\kms, respectively. In view of the masses, age, and previous history of \vstar, a standard Population I chemical composition was assumed throughout. 

We adopted surface gravity ($\log g$)
values of 2.5 and 2.0, close to our final determinations from the
analysis below.  Small adjustments typically under 1~\kms\ were
applied to the velocities to correct for systematic errors caused by
lines shifting in and out of the narrow spectral window as a function
of orbital phase \citep[see][]{Latham:1996}.  The resulting radial
velocities in the heliocentric frame including all corrections are
listed in Table~\ref{tab:rvs}, along with their uncertainties. The
light ratio between the components was determined to be $\ell_2/\ell_1
= 0.498 \pm 0.018$ at the mean wavelength of our observations.

In addition to our own, we make use below of the radial-velocity
measurements for \vstar\ by \cite{Imbert:1987}, obtained between 1977
and 1986 with the CORAVEL instrument \citep{Baranne:1979}. The projected rotational velocities \citep{Benz:1981, Benz:1984} reported by \cite{Imbert:1987} for the primary and secondary star are $17.2 \pm 0.3$~\kms\ and $21.6 \pm 0.5$~\kms, significantly more precise than ours, but consistent with them.

\subsection{Spectroscopic orbital solution}
\label{sec:specorbit}

Independent spectroscopic orbital fits using our own RVs and those of
\cite{Imbert:1987} gave results in good agreement with each other. For
the CfA measurements we detected a small, but statistically significant
systematic offset between the primary and secondary velocities of
$-0.65 \pm 0.16$~\kms\ (primary minus secondary) that may be due to
template mismatch, or perhaps the presence of spots on the presumably
more active secondary (see below).  Neither set of observations
indicated any significant eccentricity. 

For the final spectroscopic solution we combined the two RV data sets with times of minimum light from our own photometry and from the literature (Table~\ref{tab:minima} below), which help to improve the 
ephemeris. We allowed for a primary/secondary offset for 
both the CfA and CORAVEL velocities, in addition to a global shift between the two. 
Separate scale factors for the internal RV errors and for the uncertainties of the times of eclipse were iterated to achieve reduced $\chi^2$ values near unity for each component and each type of observation.
Once again, tests indicated the orbit is consistent 
with being circular.

\begin{figure}
\centering
\includegraphics[width=8.5cm]{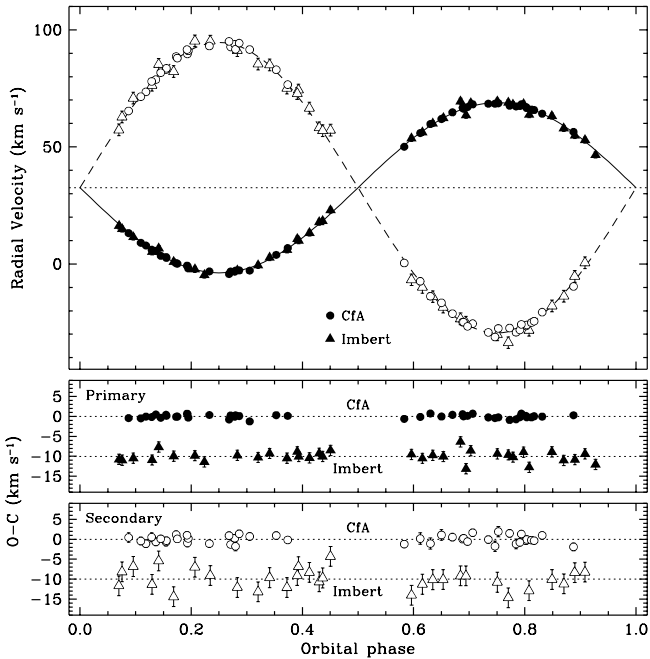}
\caption{Spectroscopic orbital solution for \vstar. Primary
  observations are shown with filled symbols; the dotted
  line in the top panel marks the center-of-mass velocity of the
  system. Residuals are shown at the bottom; the \cite{Imbert:1987} 
  residuals are shifted for clarity. \label{fig:orbit}}
\end{figure}

A weighted least-squares fit to the data gives the elements listed in
Table~\ref{tab:specorbit}. The fit and the observations may be seen in
Figure~\ref{fig:orbit}.

\input ntable2.tex

\section{Photometry}
\label{sec:lightcurve}

\subsection{Observations}
\label{sec:photobs}

\vstar\ was observed in the $uvby$ photometric system on a total of 
119 nights between 1988 and 1994 with the Danish 50\,cm Str{\"o}mgren Automatic Telescope (SAT) and its 6-channel photometer, using a 17\arcsec\ diaphragm. HD\,42131, HD\,41873, and HD\,41143 were selected as comparison stars and were observed alternately with \vstar. They were found to be constant to within 4 mmag (rms) throughout the period, and the instrument has also seen extensive use in the same period for other work. Details on the extinction corrections and transformations to the standard system are as described by \cite{Olsen:1994}. 

\setcounter{table}{3}
\input ntable4.tex

\begin{figure}[bh]
\centering
\includegraphics[width=8.5cm]{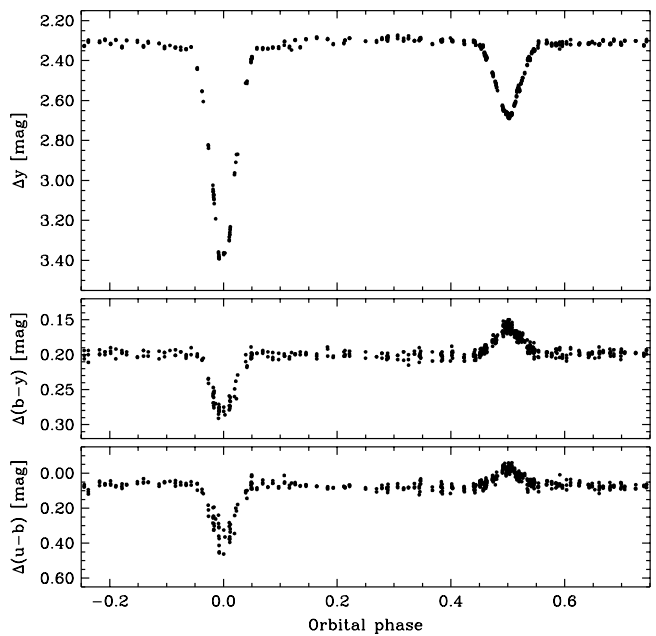}
\caption{Differential magnitudes and colors for \vstar\ on the instrumental system. \label{fig:lc}}
\end{figure}

The light curves contain a total of 559 points in each colour, with typical mean errors for a single differential magnitude of 0.008~mag ($vby$) and 0.010--0.015~mag ($u$); the $y$ observations from one season (35 observations on seven nights in 1993) fall $\sim$0.02 mag below the rest (presumably due to slight spot activity) and were simply omitted in the following analysis. All 524 remaining observations are available electronically (only) in Table~3, and are shown graphically in Figure~\ref{fig:lc}. The duration of the eclipses is about 5.5 days; Min I shows a total phase of only $\sim$16 hours. 

Standard $uvby\beta$ indices of \vstar\ at maximum light were measured relative to a grid of standard stars \citep[see][for details]{Olsen:1993, Olsen:1994}, and are $V = 9.365 \pm 0.007$, $b-y = 0.812 \pm 0.004$, $m_1 = 0.237 \pm 0.026$, and $c_1 = 0.482 \pm 0.018$ based on 10 measurements, and $\beta = 2.591$ (single measurement).

\begin{figure}
\centering
\includegraphics[width=8.5cm]{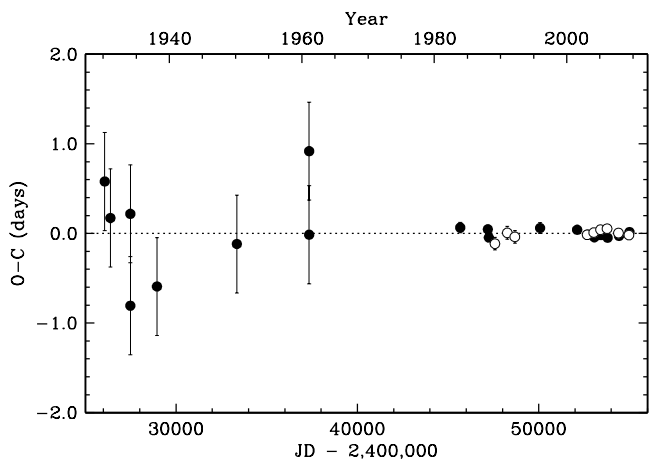}
\caption{A plot of the times of minimum listed in Table \ref{tab:minima}. 
The period seems to have been constant over 80 years.
\label{fig:period}}
\end{figure}

Times of minimum light determined by the \cite{Kwee:1956} method from groups of consecutive nights covering the eclipses are presented in Table~\ref{tab:minima}, along with other such measurements from the literature. No significant period change is detectable, and the final linear ephemeris we derive from a joint analysis with the spectroscopy (Table~\ref{tab:specorbit}) is:
\begin{equation}
\label{eph}
\mathrm{Min~I} = \mathrm{HJD}~2,\!447,\!193.554 + 52.42136\, E\,.
\end{equation}
As can be seen Figure \ref{fig:period}, the period has been constant within the limitations of the observations, contrary to the speculations of \cite{Eggleton:2017} on a possible third star in the system.

\vstar\ is well detached, and the primary eclipse, when the cooler and larger secondary is in front, has a short interval of totality (see above), so the luminosity of both stars is accurately determined. 

\subsection{Light curve solution}

We modelled the four $uvby$ light curves simultaneously using the Wilson-Devinney light curve model and program
\citep{Wilson:1971, Wilson:1979} version 2013\footnote{  \url{ftp://ftp.astro.ufl.edu/pub/wilson/lcdc2013/}.} in mode 2, called within a Markov Chain Monte Carlo (MCMC) framework in order to sample the
posterior distributions of the adjusted parameters. The main variables
we explored are the surface potentials $\Omega_1$ and $\Omega_2$, the
secondary temperature $T_{\rm eff,2}$, the inclination angle $i$, and
the out-of-eclipse magnitude $m_0$ for each passband, at phase
0.25. The primary temperature was held fixed at the spectroscopic
value $T_{\rm eff,1} = 5210$~K (Section~\ref{sec:spectroscopy}). 

We also fixed the period and reference epoch of primary eclipse to 
the values in Eq.~\ref{eph} and Table~\ref{tab:specorbit}, but we 
allowed for minor phase
adjustments $\Delta\phi$ separately for each passband. Scale factors $f_{\sigma}$
for the initial photometric uncertainties in $uvby$ ($\sigma = 0.02$~mag) were
included as free parameters as well, with the addition of appropriate
terms to the likelihood function \citep{Gregory:2005}. The mass ratio
was held at the spectroscopic value of $q = 0.5857$, and gravity
darkening exponents (0.32) and bolometric albedos (0.5) set to values
appropriate for stars with convective envelopes. To model the local emission from the two stars we used the model atmosphere option in the code. Limb darkening
coefficients $u_1$ and $u_2$ were taken from the tables of
\cite{Claret:2011} for the linear law (ATLAS models), as a more complicated treatment
is not warranted given the visible distortions in the light curves
caused by spots.

Our method of solution used the 
{\tt emcee\/}\footnote{\url{http://dan.iel.fm/emcee.}} 
code of \cite{Foreman-Mackey:2013},
which is a Python implementation of the affine-invariant MCMC ensemble
sampler proposed by \cite{Goodman:2010}. Convergence was judged by
visual inspection of the chains, along with the requirement of a
Gelman-Rubin statistic smaller than 1.05 for each parameter
\citep{Gelman:1992}.

\begin{figure*}
\centering
\includegraphics[width=18cm]{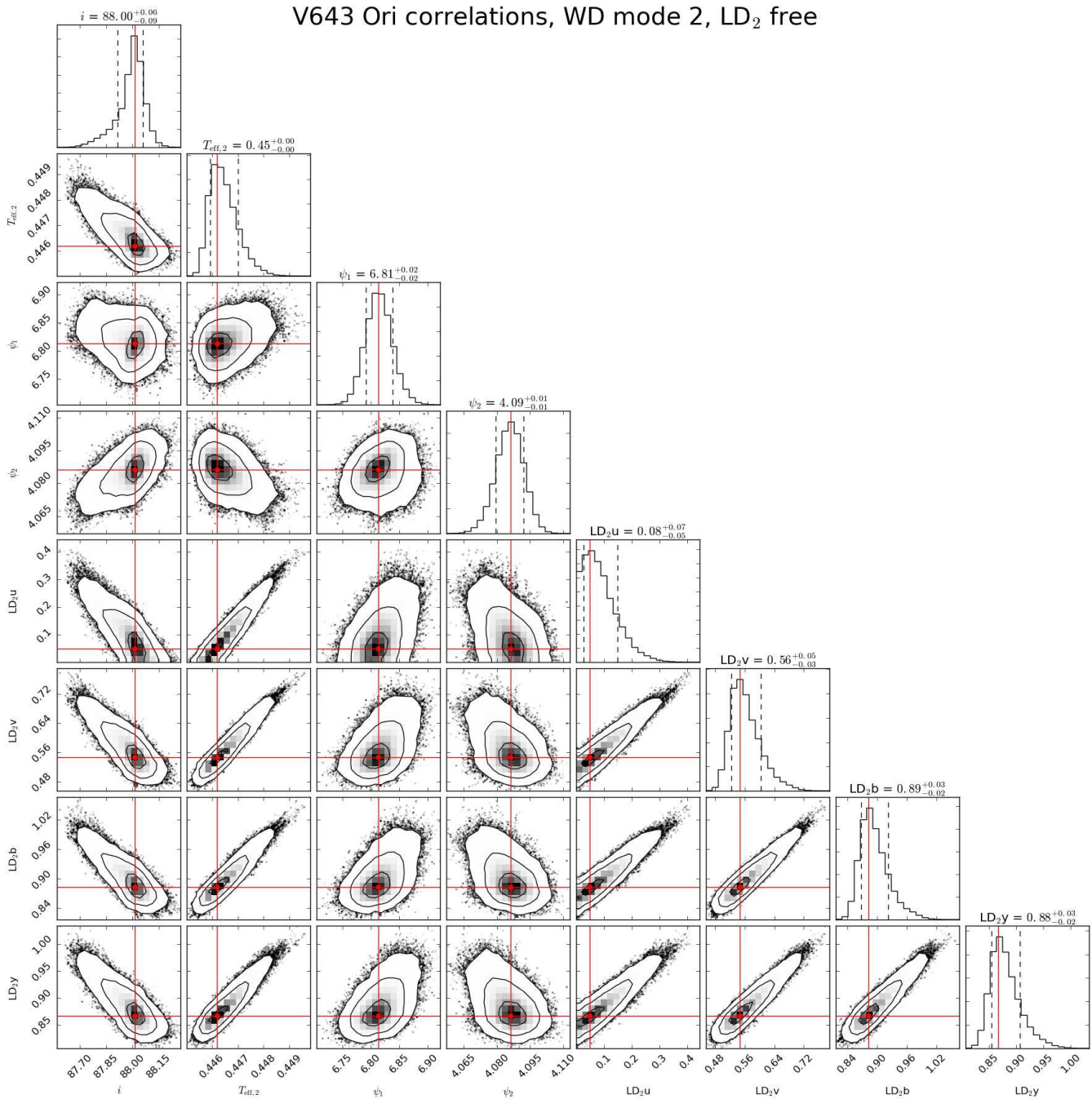}
\caption{``Corner plot'' \citep[][source code available at {\tt
      https://github.com/dfm/corner.py}]{Foreman-Mackey:2016} for
  \vstar, illustrating the correlations among the main fit parameters
  of our solution.  Contour levels correspond to 1, 2, and
  3$\sigma$, and the histograms on the diagonal represent the
  posterior distribution for each parameter, with the mode and
  internal 68\% confidence levels indicated. More realistic errors are
  discussed in the text}.\label{fig:corner} 
\end{figure*}

Initial runs resulted in an obvious systematic and
wavelength-dependent pattern to the residuals during the secondary
eclipse, likely related to the presence of spots. We found this could
be corrected by allowing the limb darkening coefficient of that star
to be a free parameter. The primary limb darkening coefficients were
held fixed, as varying them did not improve the residuals any further.
Other tests that included third light gave values of $L_3$ not
significantly different from zero that also brought no improvement, so
our final runs were carried out with $L_3 = 0$ in all bands. The
correlations among some of the key variables may be appreciated in the
diagram of Figure~\ref{fig:corner}, generated from our final solution.

\begin{figure}[th]
\centering
\includegraphics[width=8.5cm]{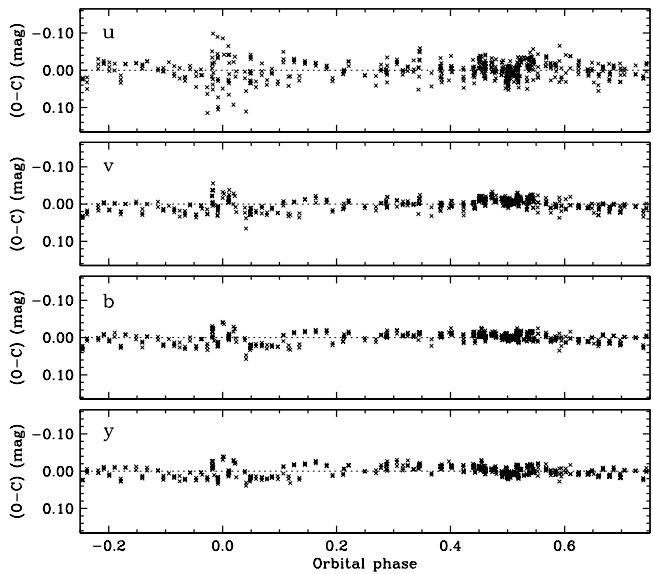}
\caption{Light curve residuals in $uvby$ from our final model for
  \vstar.\label{fig:residuals}}
\end{figure}

Time-correlated noise in the light curves as a result of the presence of spots causes the formal uncertainties of the adjusted parameters to be
underestimated. To account for this, we carried out a residual
permutation exercise in which we simulated $uvby$ data sets based on
the residuals from our adopted solution below, shifting the residuals
by a random number of time steps and adding them back into the adopted
model light curves at each time of observation. We generated 50 such
artificial data sets and submitted each one to a new MCMC analysis.
Each of these used randomly perturbed values for all previously fixed
quantities, including the primary limb darkening coefficients, the
bolometric albedos, and the gravity darkening exponents (all assumed
to be normally distributed around their adopted values, with $\sigma =
0.1$). 

\input ntable5.tex

We also applied Gaussian perturbations to the mass ratio and
primary effective temperature by amounts consistent with their
measured uncertainties. We calculated the scatter of the distributions
for each parameter, and combined those values in quadrature with the
internal errors from our nominal solution to arrive at the final
uncertainties. Table~\ref{tab:photorbit} gives the final light
curve elements and adopted uncertainties.

\section{Discussion}

\subsection{Absolute dimensions}
\label{sec:dimensions}

The preceding sections have described the analysis of our spectroscopic and photometric observations of \vstar. We now combine the computed results of Tables~\ref{tab:specorbit} and \ref{tab:photorbit} into the coherent, comprehensive and accurate picture of the physical properties of the \vstar\ system in its present state, which is summarised in Table~\ref{tab:dimensions}. 

The absolute masses and all orbital parameters are determined to about 0.5\% real accuracy and the radii to 1.5\%, or better. Both components are near-spherical and relatively far separated from their Roche lobes
(Table~\ref{tab:photorbit}) as expected, since the orbit has been circularised; \vstar\ is a well-detached system. The measured component rotations are those expected for the observed radii and the period (and inclination) of the circular orbit.

The radiative properties of \vstar\ are less precisely determined. A reddening estimate of $E(B-V) = 0.298 \pm 0.096$ was taken from the tridimensional map of the local interstellar medium of \cite{Capitanio:2017} and used to estimate the luminosities of the stars, adopting a distance to \vstar\ of 1270~pc from the {\it Gaia}/DR2 catalogue. The independent distance estimate we derive, $840 \pm 130$~pc, is somewhat smaller than the trigonometric value, possibly due to small errors in our (primary) effective temperature, or in the reddening. Thus, the present state of \vstar\ is very precisely determined, thanks both to the very favourable configuration of \vstar\ itself and to our careful analysis to obtain a consistent picture of the system, with few or no loose ends.

\input ntable6.tex

\subsection{The evolution of \vstar}
\label{sec:evolution}

Any attempt to model the evolutionary history of \vstar\ must begin by considering its initial configuration and assigning initial masses to the two stars, and a separation between them or (equivalently) the orbital period of the binary. 
The task is then to match the present configuration of \vstar\ with ($i$) a single age, ($ii$) a 52-day circular orbit, and ($iii$) two evolved, detached, synchronised components, with a physically realistic model for the evolution and mass exchange processes of the stars in \vstar. The amounts of mass and angular momentum transferred between the stars, and from the system, may be varied as necessary.

The initial primary star must obviously have been the present secondary, since it is the more highly evolved star, i.e., it must have had a mass above $\sim$2.6 $M_{\sun}$, which is half the mass of the present system.  If mass is conserved, the present secondary star must have had an initial mass substantially lower than this, since it has received some mass from the primary. Now it has become a $\sim$3.3 $M_{\sun}$ evolved star, most likely a clump giant in the core He-burning phase --- the longest-lived post-main sequence phase (both stars are detached clump giants).

\begin{figure}
\centering
\includegraphics[width=8.5cm]{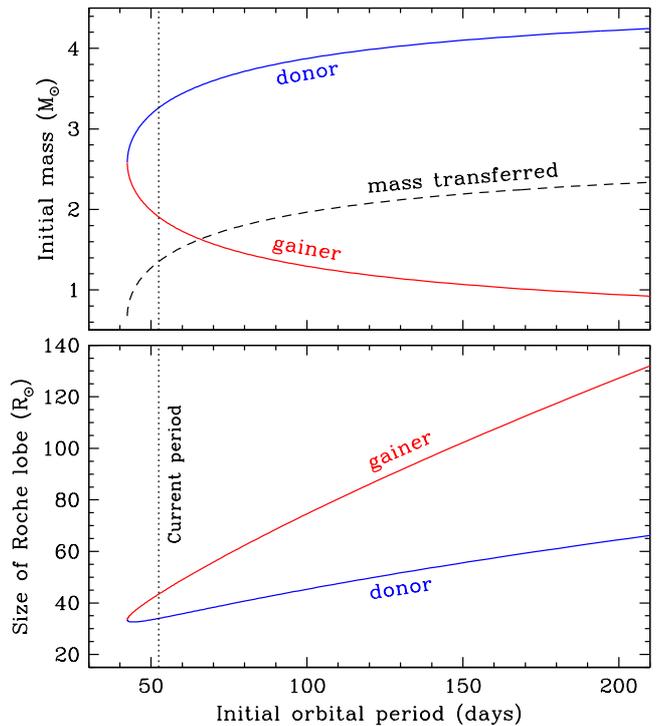}
\caption{Initial donor and gainer masses (top) required to match the present masses and orbital period of \vstar\ if total mass and angular momentum are perfectly conserved, as a function of the chosen initial period. The amount of mass transferred is also indicated. The bottom panel shows the size (radius) of the limiting Roche lobe of the donor (and gainer) at the chosen initial orbital period. Mass transfer by Roche lobe overflow begins when the donor reaches this size during its normal evolution.
\label{fig:conservative}}
\end{figure}

The simplest possible model, though arguably not the most realistic, is one in which both mass and angular momentum are perfectly conserved during mass transfer, and the orbit remains circular. Under those conditions the initial and final masses and orbital periods are related by
\begin{equation}
\frac{P_{\rm final}}{P_{\rm init}} = \left(\frac{M_{\rm 1,init} M_{\rm 2,init}}{M_{\rm 1,final} M_{\rm 2,final}}\right)^3
\label{eq:conservative}
\end{equation}
 \citep[see, e.g.][]{Postnov:2014}. With the final state known, and the initial secondary mass given by $M_{\rm 2,init} = M_{\rm tot} - M_{\rm 1,init}$, any choice for the initial orbital period immediately determines the initial primary mass. We illustrate this in Figure~\ref{fig:conservative}.

One possible scenario has the \vstar\ system starting out with the present orbital period of 52 days, and with the initial masses reversed from what we measure. The size of the Roche lobe of the original 3.3~$M_{\sun}$ donor would be roughly 34~$R_{\sun}$. A solar-metallicity evolutionary track calculated with the MESA code \citep{Paxton:2011, Paxton:2013, Paxton:2015} shows that a normal star of that mass would reach this size near the tip of the giant branch (first ascent), at an age of about 290~Myr, triggering ``Case~B'' mass transfer by Roche lobe overflow (RLOF). Conservation of angular momentum dictates that the orbit must shrink as mass flows from the donor to the less massive gainer until the masses become equal.

With the equation above we calculate that at this point the orbital period reaches a minimum of about 42 days, implying a separation near 90~$R_{\sun}$. The $\sim$33~$R_{\sun}$ Roche lobes of the two stars are thus well separated, consistent with the notion of conservation of mass. 

As the transfer of material in this idealized scenario continues, the orbit widens again, and mass exchange ceases when the donor retreats as a core helium burning (CHB or `clump') giant inside its $\sim$44~$R_{\sun}$ Roche lobe at the present 52-day period. 

By then the former secondary has accreted about 1.4~$M_{\sun}$ worth of Population~I material, and ends up appearing as a fairly typical 3.3~$M_{\sun}$ clump giant. Comparison of its measured properties with those predicted from a MESA evolutionary track shows it to have the right luminosity for a clump giant, while being slightly smaller and hotter than expected. On the other hand the original primary star, now less massive, but also a CHB giant with a $\sim0.6 M_{\sun}$ He core, is overluminous, large, and cool for its mass, compared to a normal clump giant, but still fitting comfortably inside its Roche lobe.

While matching many of the present properties of the system, the configuration explored above, with an initial orbital period that is the same as the current one, is of course completely arbitrary and many other choices of initial configuration are possible, according to Figure~\ref{fig:conservative}, all leading to the same final masses and period.

A somewhat more realistic scenario would account for the possibility of mass and angular momentum loss from the system. This may occur, e.g., if a fraction $\beta$ of the mass that reaches the vicinity of the gainer is then ejected as a fast, isotropic wind with the specific angular momentum of the gainer. Under those conditions the relationship between the initial and final periods becomes
\begin{equation}
\frac{P_{\rm final}}{P_{\rm init}} = \left(\frac{q_{\rm init}}{q_{\rm final}}\right)^3 \left(\frac{1+q_{\rm init}}{1+q_{\rm final}}\right)^2 \left(\frac{1+(1-\beta)q_{\rm final}}{1+(1-\beta)q_{\rm inital}}\right)^{\frac{8-5\beta}{1-\beta}}
\label{eq:soberman}
\end{equation}
\citep[see, e.g.,][]{Tauris:1996, Soberman:1997}, in which the mass ratio is defined as $q \equiv M_1/M_2$ and star~1 is the donor. The orbit is assumed to remain circular throughout. 

\begin{figure}
\centering
\includegraphics[width=8.5cm]{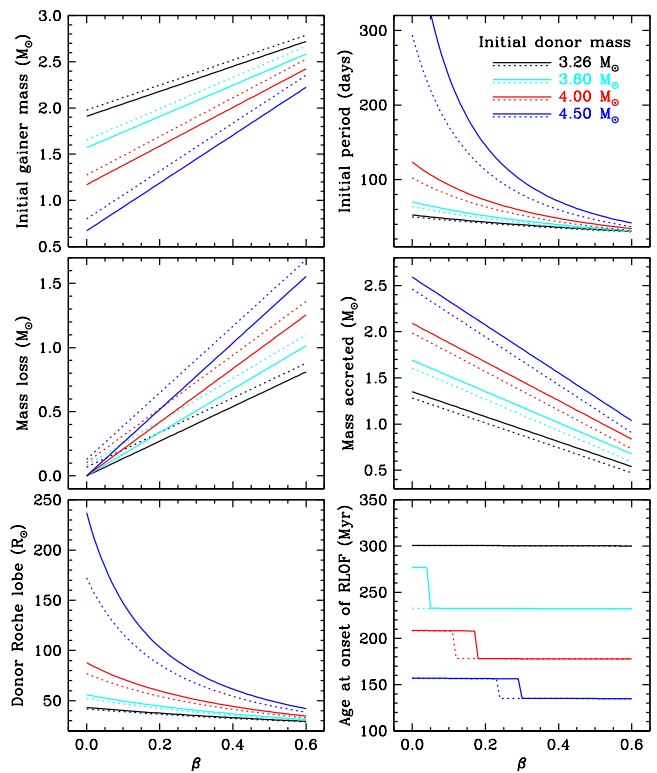}
\caption{Initial conditions and other properties of the \vstar\ system under non-conservative mass transfer, calculated with the formalism of \cite{Soberman:1997}. Results are plotted for four initial donor masses (labeled), as a function of the fraction $\beta$ of the mass transferred from the donor that is ejected from the system in the vicinity of the gainer. The middle panels represent the amount of mass lost, and the amount of material that is actually accreted by the gainer. The bottom panels give the size of the donor's Roche lobe at the initial period (left), and the age of the system (right) when the donor triggers mass transfer as it reaches that size, according to MESA models for normal solar-metallicity stars. The discontinuities in the age sequences as $\beta$ increases are the result of the initial donor switching from attaining the critical (Roche-lobe) radius during the asymptotic giant branch phase (larger sizes) to reaching it on the first ascent of the giant branch (smaller radii, hence younger ages). Dotted lines in all panels show how the results change when an additional 5\% of the mass transferred is lost to the system via an isotropic wind from the donor star (see text).
\label{fig:soberman}}
\end{figure}

For a given initial donor mass, Figure~\ref{fig:soberman} shows the initial gainer mass and initial orbital period that are required in order to reach the present masses and period after the end of RLOF. This is given as a function of the fraction $\beta$ of mass transferred that is lost from the system in the manner described above. 

Other properties of the system resulting from this non-conservative scenario may also be seen in the figure, including the initial radius of the donor's Roche lobe, and the age of the star when it evolves to reach that size, and mass transfer begins. When $\beta = 0$, Eq.~\ref{eq:soberman} reduces to Eq.~\ref{eq:conservative} and the results are exactly the same as in the strictly conservative scenario considered earlier. The non-conservative scenario has one more free parameter than the conservative case, and as before, there are many possible sets of initial conditions that lead to the current configuration.

The \cite{Soberman:1997} formalism is general enough that other mechanisms of mass and angular momentum loss may be explored as well. One is a spherically symmetric outflow from the donor star in the form of a fast, magnetized wind, carrying with it the specific angular momentum of the donor --- not unlikely in view of the observed spot(s) presumably on the present secondary (cooler star). Another is mass loss to a circumbinary accretion ring. As an example, the first of these mechanisms is illustrated in Figure~\ref{fig:soberman} by the dotted lines, calculated for a loss of 5\% of the transferred mass \citep[parameter $\alpha$ in the notation of][]{Soberman:1997} through an isotropic wind. The effect is quite significant, especially at the higher initial donor masses.

Unfortunately, these increasingly complex scenarios add more free parameters but bring us no closer to a unique solution for \vstar. They give us no detailed information about what transpires during mass transfer, which will surely result in very different stellar structures and final sizes and temperatures for the two stars (both of which are well determined) for any given choice of initial conditions. 

Furthermore, these scenarios may still be too simplistic; more detailed modelling would be desirable to account for other forms of mass and angular momentum loss from the system that may be at work, as well as for the evolution of the initial orbital eccentricity and rotation rates due to tides (since we know that the present system is circular and synchronised), and even for a possible common-envelope phase. This should allow important additional constraints such as the measured radii and temperatures of the components to narrow the range of possible solutions, ideally down to a single one.

At our request, attempts in that direction were carried out by J.\ Andrews using the BSE code \citep{Hurley:2002} and also by A.\ Dotter with the more sophisticated MESA binary code \citep{Paxton:2015}. Both computer programmes make tradeoffs and have their own limitations: BSE is able to follow the binary system through the common-envelope phase in an approximate way, but makes many simplifications in dealing with evolution; MESA treats stellar evolution more rigorously, but cannot handle the common-envelope phase. As a result, neither experiment succeeded in identifying a plausible initial state leading to the current properties of the system. 

With BSE, a wide range of initial properties leads to the system entering a common-envelope phase, from which it emerges with masses and/or an orbital period very different from what we measure. The attempts to model the system with MESA, on the other hand, seem to fail because it is not designed for systems in which mass transfer begins when the donor is already a giant, as is likely the case for \vstar. In the end, therefore, these efforts were less satisfactory than the simpler calculations described above.

\section{Conclusion}
\label{sec:conclusion}

Our original goal of properly modelling in detail the evolution of \vstar, taking into account the exchange of mass and angular momentum and their loss from the stars and from the system itself, thus eluded us, even though the end product --- the present system --- was so precisely specified. At the present time, theoretical models (especially for convective giant stars) seem inadequate to treat these processes in sufficient detail to allow the use of all observational constraints at the same time, and in that way to identify a unique solution. 

In the meantime, we have at least explored the variation of the masses and orbital period under conservative and non-conservative scenarios, though without following the evolution of other stellar properties in any detail. When full models allowing this finally do become available, we shall have at least one carefully observed system and its properties to look for!

\begin{acknowledgements}

Service observers J.\ Caruso, R.\ Davis, R.\ Stefanik, and J.\ Zajac helped with the spectroscopic observations, and several SAT observers included \vstar\ in their programmes; we thank them all. Jeff Andrews and Aaron Dotter spent considerable time in trying to model the evolution of \vstar\ with MESA and other suites of evolutionary codes, and P.\ Maxted helped early on with the determination of the stellar properties.  JVC participated fully in the observations and their reduction, so he clearly merits posthumous co-authorship, but he obviously bears no responsibility for the analysis and final text of the paper.\\
We thank the careful anonymous referee for providing very helpful comments on the original manuscript, which led to extensive modifications.
GT acknowledges partial support from the NSF through grant AST-1509375, and the use of the Danish telescopes on La Silla was supported by the then Danish Natural Science Research Council (now the Independent Research Fund Denmark | Natural Sciences). This research has made use of the SIMBAD database, operated at the CDS, Strasbourg, France, and of NASA's Astrophysics Data System Abstract Service.

\end{acknowledgements}

\end{document}

%% file: ntable1.tex
\setlength{\tabcolsep}{2pt} 
\begin{table}
\caption{Heliocentric radial velocities of \vstar\ from CfA.}
\label{tab:rvs}
\centering
\begin{tabular}{lrcrcc}
\hline\hline \\ [-2ex]
~~~~HJD & $RV_1$ & $\sigma_1$ & $RV_2$ & $\sigma_2$ & Orbital \\
2,400,000$+$ & (\kms) & (\kms) & (\kms) & (\kms) & phase \\ [0.5ex]
\hline \\ [-2ex]
    47100.9101  &   $-$3.18  &    0.44  &     93.73  &    0.93  &   0.2326 \\
    47461.8790  &      7.88  &    0.45  &     74.14  &    0.95  &   0.1185 \\
    47462.8107  &      5.61  &    0.46  &     79.39  &    0.98  &   0.1362 \\
    47463.7724  &      2.96  &    0.49  &     83.71  &    1.03  &   0.1546 \\
    47463.8096  &      2.61  &    0.43  &     84.22  &    0.92  &   0.1553 \\
    47465.7904  &   $-$0.87  &    0.45  &     90.36  &    0.95  &   0.1931 \\
    47465.8673  &   $-$1.88  &    0.38  &     91.68  &    0.81  &   0.1946 \\
    47469.8568  &   $-$3.21  &    0.36  &     93.35  &    0.76  &   0.2707 \\
    47487.7424  &     55.91  &    0.70  &   $-$6.75  &    1.47  &   0.6118 \\
    47488.6968  &     59.78  &    0.67  &  $-$13.14  &    1.43  &   0.6301 \\
    47489.7456  &     61.94  &    0.63  &  $-$15.93  &    1.33  &   0.6501 \\
    47490.7798  &     64.76  &    0.46  &  $-$20.57  &    0.97  &   0.6698 \\
    47491.7388  &     66.71  &    0.39  &  $-$24.16  &    0.83  &   0.6881 \\
    47491.8272  &     66.32  &    0.41  &  $-$24.17  &    0.88  &   0.6898 \\
    47492.6853  &     68.18  &    0.41  &  $-$24.82  &    0.88  &   0.7061 \\
    47494.7860  &     68.44  &    0.61  &  $-$30.53  &    1.30  &   0.7462 \\
    47496.7754  &     67.29  &    0.59  &  $-$28.54  &    1.26  &   0.7842 \\
    47497.7134  &     67.04  &    0.43  &  $-$25.84  &    0.91  &   0.8021 \\
    47497.7826  &     66.61  &    0.43  &  $-$25.37  &    0.92  &   0.8034 \\
    47512.6629  &     13.19  &    0.57  &     65.99  &    1.20  &   0.0872 \\
    47513.8190  &      9.03  &    0.47  &     72.10  &    1.00  &   0.1093 \\
    47514.8394  &      6.11  &    0.54  &     78.62  &    1.13  &   0.1288 \\
    47515.6919  &      3.50  &    0.42  &     82.34  &    0.88  &   0.1450 \\
    47522.6724  &   $-$3.26  &    0.45  &     94.47  &    0.94  &   0.2782 \\
    47522.7456  &   $-$2.83  &    0.54  &     92.35  &    1.13  &   0.2796 \\
    47526.5811  &      3.89  &    0.42  &     83.62  &    0.90  &   0.3527 \\
    47527.6685  &      6.80  &    0.41  &     77.24  &    0.88  &   0.3735 \\
    47538.6520  &     50.06  &    0.44  &      1.06  &    0.94  &   0.5830 \\
    47544.6200  &     67.01  &    0.39  &  $-$25.99  &    0.83  &   0.6969 \\
    47546.5788  &     68.44  &    0.41  &  $-$28.57  &    0.87  &   0.7342 \\
    47547.5260  &     68.63  &    0.63  &  $-$26.86  &    1.33  &   0.7523 \\
    47548.5882  &     67.61  &    0.43  &  $-$26.69  &    0.92  &   0.7726 \\
    47549.5577  &     67.47  &    0.44  &  $-$27.52  &    0.93  &   0.7910 \\
    47549.6979  &     68.19  &    0.44  &  $-$25.14  &    0.93  &   0.7937 \\
    47550.6414  &     65.92  &    0.37  &  $-$24.40  &    0.79  &   0.8117 \\
    47550.8994  &     65.75  &    0.37  &  $-$23.77  &    0.79  &   0.8166 \\
    47551.6600  &     64.19  &    0.43  &  $-$19.91  &    0.92  &   0.8311 \\
    47554.6145  &     56.40  &    0.46  &   $-$8.93  &    0.97  &   0.8875 \\
    47569.5650  &      0.34  &    0.40  &     89.24  &    0.86  &   0.1727 \\
    47569.6841  &      0.16  &    0.45  &     88.55  &    0.95  &   0.1750 \\
    47570.5943  &   $-$0.74  &    0.41  &     92.20  &    0.86  &   0.1923 \\
    47574.5598  &   $-$4.27  &    0.45  &     95.70  &    0.95  &   0.2680 \\
    47575.5136  &   $-$2.71  &    0.44  &     94.96  &    0.94  &   0.2862 \\
    47576.5013  &   $-$2.83  &    0.44  &     92.21  &    0.93  &   0.3050 \\ [0.5ex]
\hline
\end{tabular}
\tablefoot{Orbital phases are computed from Eq.~\ref{eph} below.}
\end{table}
\setlength{\tabcolsep}{6pt}  

%% file: ntable2.tex
\setlength{\tabcolsep}{2pt} 
\begin{table}
\caption{Spectroscopic orbital solution for \vstar.}
\label{tab:specorbit}
\centering
\begin{tabular}{lc}
\hline\hline \\ [-2ex]
~~~~~~~~Parameter~~~~~~~~ & Value \\ [0.5ex]
\hline \\ [-2ex]
\multicolumn{2}{c}{Adjusted elements} \\
$P$ (days)                                &   52.42136~$\pm$~0.00014 \\
$\gamma$ (\kms)                           &   +32.557~$\pm$~0.067 \\
$K_1$ (\kms)                              &   36.307~$\pm$~0.073 \\
$K_2$ (\kms)                              &   61.99~$\pm$~0.16 \\
Min~I (HJD $2,\!400,\!000+$)              &   47,193.554~$\pm$~0.016 \\
$\Delta_{\rm CfA}$ (\kms)                 &   $-$0.64~$\pm$~0.16 \\
$\Delta_{\rm CORAVEL}$ (\kms)             &   $-$0.40~$\pm$~0.51 \\
$\Delta$ (\kms)                           &   $-$0.84~$\pm$~0.23 \\ 
\multicolumn{2}{c}{Derived quantities} \\
$M_1 \sin^3 i$ ($\mathcal{M}_{\sun}^{\rm N}$)               &   3.254~$\pm$~0.019 \\
$M_2 \sin^3 i$ ($\mathcal{M}_{\sun}^{\rm N}$)               &   1.9056~$\pm$~0.0090 \\
$q \equiv M_2/M_1$                        &   0.5857~$\pm$~0.0019 \\
$a \sin i$ ($\mathcal{R}_{\sun}^{\rm N}$)                   &   101.86~$\pm$~0.18 \\ [2ex]
CfA $\sigma_1$, $\sigma_2$ (\kms)         &   0.44 / 0.95 \\
CfA $N_{\rm obs,1}$, $N_{\rm obs,2}$      &   44 / 44 \\
CORAVEL $\sigma_1$, $\sigma_2$ (\kms)     &   1.31 / 2.54 \\
CORAVEL $N_{\rm obs,1}$, $N_{\rm obs,2}$  &   35 / 31 \\
$N_{\rm Min~I}$, $N_{\rm Min~II}$         &   19 / 9 \\ [0.5ex]
\hline
\end{tabular}

\tablefoot{$\Delta_{\rm CfA}$ and $\Delta_{\rm CORAVEL}$ are the
  primary minus secondary velocity offsets, $\Delta$ is the global
  CfA minus CORAVEL shift. The minimum masses and semimajor axis are
  in units of the solar mass and radius \citep[2015 IAU Resolution B3, see][]{Prsa:2016}.}

\end{table}
\setlength{\tabcolsep}{6pt}  

%% file: ntable4.tex
\setlength{\tabcolsep}{4pt} 
\begin{table}[t!]
\caption{Times of minimum light for \vstar.}
\label{tab:minima}
\centering
\begin{tabular}{lccccc}
\hline\hline \\ [-2ex]
~~~~HJD & $\sigma$ & & & & \\
(2,400,000+) & (days) & Eclipse & Type & Year & Source \\ [0.5ex]
\hline \\ [-2ex]
   26068.325    &   0.530  & 1 &   pg  &  1930.2487  &  1 \\
   26382.446    &   0.530  & 1 &   pg  &  1931.1087  &  1 \\
   27482.315    &   0.530  & 1 &   pg  &  1934.1200  &  1 \\
   27483.341    &   0.530  & 1 &   pg  &  1934.1228  &  1 \\
   28950.328    &   0.530  & 1 &   pg  &  1938.1392  &  1 \\
   33354.197    &   0.530  & 1 &   pg  &  1950.1963  &  2 \\
   37338.324    &   0.530  & 1 &   pg  &  1961.1042  &  2 \\
   37339.257    &   0.530  & 1 &   pg  &  1961.1068  &  2 \\
   45673.400    &   0.057  & 1 &   v   &  1983.9244  &  2 \\
   47193.600    &   0.046  & 1 &   pe  &  1988.0865  &  3 \\
   47245.930    &   0.046  & 1 &   pe  &  1988.2298  &  3 \\
   47586.600    &   0.074  & 2 &   pe  &  1989.1625  &  3 \\
   48268.200    &   0.074  & 2 &   pe  &  1991.0286  &  3 \\
   48687.525    &   0.074  & 2 &   pe  &  1992.1767  &  3 \\
   50076.79     &   0.057  & 1 &   v   &  1995.9803  &  4 \\
   52121.20335  &   0.031  & 1 &   pe  &  2001.5776  &  5 \\
   52671.57191  &   0.031  & 2 &   pe  &  2003.0844  &  5 \\
   52697.78540  &   0.031  & 1 &   pe  &  2003.1562  &  5 \\
   53038.54694  &   0.031  & 2 &   pe  &  2004.0891  &  5 \\
   53064.70369  &   0.031  & 1 &   pe  &  2004.1607  &  5 \\
   53405.52748  &   0.031  & 2 &   pe  &  2005.0938  &  5 \\
   53431.67806  &   0.031  & 1 &   pe  &  2005.1654  &  5 \\
   53772.48842  &   0.031  & 2 &   pe  &  2006.0985  &  5 \\
   53798.59786  &   0.031  & 1 &   pe  &  2006.1700  &  5 \\
   54401.49642  &   0.031  & 2 &   pe  &  2007.8207  &  5 \\
   54427.67544  &   0.031  & 1 &   pe  &  2007.8923  &  5 \\
   54978.10783  &   0.031  & 2 &   pe  &  2009.3993  &  5 \\
   55004.35238  &   0.031  & 1 &   pe  &  2009.4712  &  5 \\ [0.5ex]
\hline
\end{tabular}
\tablefoot{``Eclipse'' is 1 for primary minimum, 2 for secondary. ``Type'' indicates photographic (pg), visual (v),
  or photoelectric (pe). Sources are:
(1) \cite{Mauder:1960};
(2) Online Bundesdeutsche Arbeitsgemeinschaft f\"ur Ver\"anderliche
  Sterne (BAV)
  database\footnote{\url{http://var2.astro.cz/ocgate/index.php?lang=en}};
(3) This paper;
(4) \cite{Hubscher:1997};
(5) \cite{Zasche:2011}.
}
\end{table}
\setlength{\tabcolsep}{6pt}  

%% file: ntable5.tex
\setlength{\tabcolsep}{4pt} 
\begin{table*}
\caption{Adopted light curve solution for \vstar.}
\label{tab:photorbit}
\centering
\begin{tabular}{l c c c c}
\hline\hline \\ [-2ex]
Parameter  & \multicolumn{2}{c}{Primary} & \multicolumn{2}{c}{Secondary} \\ [0.5ex]
\hline \\ [-1.5ex]
$\Omega$          &  \multicolumn{2}{c}{$6.813^{+0.095}_{-0.094}$}    &  \multicolumn{2}{c}{$4.085^{+0.023}_{-0.023}$}    \\ [0.5ex]
$r_{\rm pole}$    &  \multicolumn{2}{c}{$0.1604 \pm 0.0024$}          &  \multicolumn{2}{c}{$0.2021 \pm 0.0015$}          \\ 
$r_{\rm point}$   &  \multicolumn{2}{c}{$0.1616 \pm 0.0025$}          &  \multicolumn{2}{c}{$0.2101 \pm 0.0018$}          \\ 
$r_{\rm side}$    &  \multicolumn{2}{c}{$0.1609 \pm 0.0024$}          &  \multicolumn{2}{c}{$0.2044 \pm 0.0016$}          \\ 
$r_{\rm back}$    &  \multicolumn{2}{c}{$0.1615 \pm 0.0025$}          &  \multicolumn{2}{c}{$0.2086 \pm 0.0017$}          \\ 
$r_{\rm vol}$     &  \multicolumn{2}{c}{$0.1610 \pm 0.0024$}          &  \multicolumn{2}{c}{$0.2052 \pm 0.0016$}          \\ 
$r_{\rm Roche}$   &  \multicolumn{2}{c}{0.429}                        &  \multicolumn{2}{c}{0.330}                        \\ 
\hline \\ [-1.5ex]
$R_2/R_1$            &       \multicolumn{4}{c}{$1.275 \pm 0.018$}         \\ [0.5ex]
$i$ (deg)            &       \multicolumn{4}{c}{$88.02^{+0.34}_{-0.35}$}   \\ [1ex]
$T_{\rm eff,1}$ (K)  &       \multicolumn{4}{c}{5210 (fixed)}              \\ [0.5ex]
$T_{\rm eff,2}$ (K)  &       \multicolumn{4}{c}{$4462^{+96}_{-95}$}        \\ [0.5ex]
$\Delta T_{\rm eff}$ (K) & \multicolumn{4}{c}{$748 \pm 51$}                \\ [0.5ex]
\hline \\ [-2ex]
Parameter  &                       $u$                 &               $v$                  &                 $b$                &                $y$                \\ [0.5ex]
\hline \\ [-2ex]
$u_1$ (fixed)    &                0.939                &              0.922                 &                0.795               &              0.674                \\ [0.5ex]
$u_2$            &   $0.05^{+0.21}_{-0.05}$            &  $0.54^{+0.13}_{-0.12}$            &  $0.88^{+0.11}_{-0.10}$            &  $0.87^{+0.12}_{-0.12}$           \\ [1ex]
$\ell_2/\ell_1$  &   $0.370 \pm 0.010$                 &  $0.445 \pm 0.009$                 &  $0.546 \pm 0.010$                 &  $0.616 \pm 0.012$                \\ [0.5ex]
$\Delta\phi$     &   $-0.00045^{+0.00017}_{-0.00017}$  &  $-0.00067^{+0.00015}_{-0.00014}$  &  $-0.00061^{+0.00016}_{-0.00015}$  &  $-0.00055^{+0.00016}_{-0.00015}$ \\ [1ex]
$m_0$ (mag)      &   $2.5768^{+0.0034}_{-0.0034}$      &  $2.5315^{+0.0031}_{-0.0031}$      &  $2.4992^{+0.0033}_{-0.0032}$      &  $2.3027^{+0.0025}_{-0.0025}$     \\ [1ex]
$f_{\sigma}$     &   $1.297^{+0.052}_{-0.043}$         &  $0.756^{+0.042}_{-0.039}$         &  $0.672^{+0.039}_{-0.036}$         &  $0.671^{+0.042}_{-0.041}$        \\ [1ex]
$\sigma$ (mag)   &               0.0260                &             0.0151                 &               0.0135               &            0.0134                 \\ [0.5ex]
\hline
\end{tabular}
\end{table*}
\setlength{\tabcolsep}{6pt}  

%% file: ntable6.tex
\setlength{\tabcolsep}{4pt} 
\begin{table}
\caption{Physical Properties of \vstar.}
\label{tab:dimensions}
\centering
\begin{tabular}{lcc}
\hline\hline \\ [-2ex]
~~~~~~~Parameter~~~~~~~ & Primary & Secondary \\ [0.5ex]
\hline \\ [-2ex]
 $M$ ($\mathcal{M}_{\sun}^{\rm N}$)  &  $3.260 \pm 0.019$  &  $1.9090 \pm 0.0091$  \\ 
 $R$ ($\mathcal{R}_{\sun}^{\rm N}$)  &  $16.41 \pm 0.25$   &  $20.91 \pm 0.17$     \\ 
 $a$ ($\mathcal{R}_{\sun}^{\rm N}$)  &   \multicolumn{2}{c}{$101.92 \pm 0.18$}     \\ 
 $\log g$ (dex)                      &  $2.521 \pm 0.013$  &  $2.0782 \pm 0.0072$  \\ 
 $T_{\rm eff}$ (K)                   &  5210~$\pm$~150     &  4460~$\pm$~180       \\ 
 $\log L/L_{\sun}$                   &  $2.252 \pm 0.052$  &  $2.194 \pm 0.070$    \\ 
 $M_{\rm bol}$ (mag)                 &  $-0.90 \pm 0.13$   &  $-0.75 \pm 0.17$     \\ 
 $BC_V$ (mag)                        &  $-0.22 \pm 0.11$   &  $-0.65 \pm 0.18$     \\ 
 $M_V$ (mag)                         &  $-0.68 \pm 0.21$   &  $-0.10 \pm 0.33$     \\ 
 $v_{\rm sync} \sin i$ (\kms)\tablefootmark{a}  &  $15.83 \pm 0.24$  &  $20.27 \pm 0.26$ \\ 
 $v \sin i$ (\kms)\tablefootmark{b}  &  $17.2 \pm 0.3$     &  $21.6 \pm 0.5$       \\ 
 $E(B-V)$ (mag)                      &   \multicolumn{2}{c}{0.298~$\pm$~0.093}     \\ 
 $A_V$ (mag)                         &   \multicolumn{2}{c}{0.92~$\pm$~0.29}       \\ 
 Dist.\ modulus (mag)                &   \multicolumn{2}{c}{$9.62 \pm 0.34$}       \\ 
 Distance (pc)                       &   \multicolumn{2}{c}{$840 \pm 130$}         \\ 
 $\pi$ (mas)                         &   \multicolumn{2}{c}{$1.19 \pm 0.19$}       \\ 
 $\pi_{Gaia/{\rm DR2}}$ (mas)        &   \multicolumn{2}{c}{$0.786 \pm 0.048$} \\ [0.5ex]
\hline
\end{tabular}
\tablefoot{The masses, radii, and semimajor axis $a$ are expressed
  in units of the nominal solar mass and radius
  ($\mathcal{M}_{\sun}^{\rm N}$, $\mathcal{R}_{\sun}^{\rm N}$) as
  recommended by 2015 IAU Resolution B3 \citep[see][]{Prsa:2016}, and
  the adopted solar temperature is 5772~K (2015 IAU Resolution
  B2). Bolometric corrections are from the work of \cite{Flower:1996},
  with conservative uncertainties of 0.1~mag added in quadrature, and
  the bolometric magnitude adopted for the Sun appropriate for this
  $BC_V$ scale is $M_{\rm bol}^{\sun} = 4.732$
  \citep[see][]{Torres:2010b}. See text for the source of the
  reddening. For the apparent visual magnitude of \vstar\ out of
  eclipse we used $V = 9.365 \pm 0.020$ (Section~\ref{sec:photobs}, with a more conservative error).
\tablefoottext{a}{Synchronous projected rotational velocity assuming
  spin-orbit alignment.}
\tablefoottext{b}{Measured projected rotational velocity \citep{Imbert:1987}.}}
\end{table}
\setlength{\tabcolsep}{6pt}  